\documentclass[sigconf,nonacm]{acmart}

\usepackage{booktabs}
\usepackage{graphicx}
\usepackage{listings}
\usepackage{xcolor}
\usepackage{enumitem}
\usepackage{microtype}
\usepackage{float}
\usepackage{xspace}
\usepackage{pifont}
\usepackage{makecell}
\usepackage{colortbl}

\newcommand{\system}{\textsc{Sovereign-OS}\xspace}
\newcommand{\cmark}{{\color{green!60!black}\ding{51}}}
\newcommand{\xmark}{{\color{red!70!black}\ding{55}}}

\renewcommand{\texttt}[1]{%
  \begingroup\ttfamily\hyphenchar\font=45
  \spaceskip=.5em plus .3em minus .2em
  #1\endgroup}


\renewcommand\footnotetextcopyrightpermission[1]{}
\AtBeginDocument{\pagestyle{plain}}

\begin{document}

\acmConference{}{}{}
\authorsaddresses{}

\title{Sovereign-OS: A Charter-Governed Operating System for Autonomous AI Agents with Verifiable Fiscal Discipline}

\author{\mbox{Aojie Yuan\textsuperscript{1}}\qquad \mbox{Haiyue Zhang\textsuperscript{1}}\qquad \mbox{Ziyi Wang\textsuperscript{2}}\qquad \mbox{Yue Zhao\textsuperscript{1}}}
\affiliation{%
  \institution{\textsuperscript{1}University of Southern California\qquad \textsuperscript{2}University of Maryland}
  \country{}
}
\email{aojieyua@usc.edu, haiyuez@usc.edu, zoewang@umd.edu, yue.z@usc.edu}

\begin{abstract}

As AI agents evolve from text generators into autonomous economic actors that accept jobs, manage budgets, and delegate to sub-agents, the absence of runtime governance becomes a critical gap. Existing frameworks orchestrate agent behavior but impose no fiscal constraints, require no earned permissions, and offer no tamper-evident audit trail.

We introduce \system{}\footnote{Open-source (MIT). \textbf{Code}: \url{https://github.com/Justin0504/Sovereign-OS}. \\ \textbf{Demo video}: \url{https://youtu.be/vOarAI_epb4}}, a governance-first operating system that places every agent action under constitutional control. A declarative \emph{Charter} (YAML) defines mission scope, fiscal boundaries, and success criteria. A \emph{CEO} (Strategist) decomposes goals into dependency-aware task DAGs; a \emph{CFO} (Treasury) gates each expenditure against budget caps, daily burn limits, and profitability floors via an auction-based bidding engine; \emph{Workers} operate under earned-autonomy permissions governed by a dynamic TrustScore; and an \emph{Auditor} (ReviewEngine) verifies outputs against Charter KPIs, sealing each report with a SHA-256 proof hash.

Across our evaluation suite, \system blocks 100\% of fiscal violations (30 scenarios), achieves 94\% correct permission gating (200 trust-escalation missions), and maintains zero integrity failures over 1{,}200+ audit reports. The system further integrates Stripe for real-world payment processing, closing the loop from task planning to revenue collection.

Our live demonstration walks through three scenarios: loading distinct Charters to observe divergent agent behavior, triggering CFO fiscal denials under budget and profitability constraints, and escalating a new worker's TrustScore from restricted to fully authorized with on-the-spot cryptographic audit verification.

\end{abstract}


\maketitle

\begin{figure*}[t]
    \centering
    \includegraphics[width=0.93\textwidth]{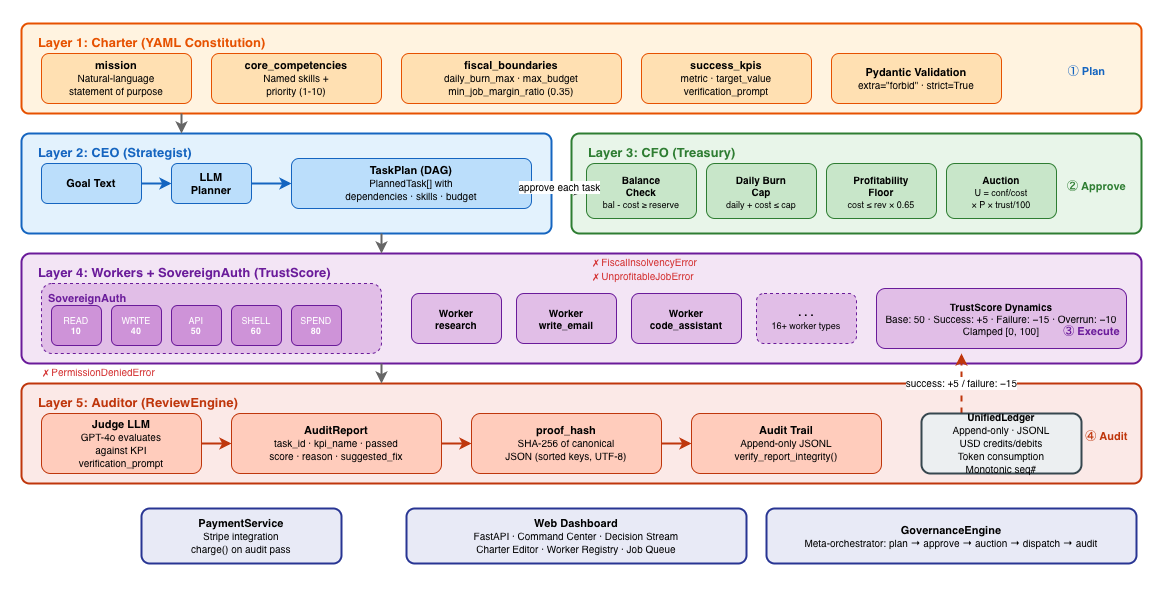}
    \vspace{-0.1in}
    \caption{\system five-layer architecture. The Charter defines mission scope, fiscal boundaries, and success KPIs. The CEO (Strategist) decomposes goals into task DAGs mapped to Charter competencies. The CFO (Treasury) enforces budget approval, daily burn caps, and job profitability floors. Workers execute tasks subject to TrustScore-gated permissions (SovereignAuth). The Auditor (ReviewEngine) evaluates outputs against KPI verification prompts and produces SHA-256-hashed AuditReports. An append-only UnifiedLedger records all financial and token flows.}
    \vspace{-0.1in}
    \label{fig:architecture}
\end{figure*}

\section{Introduction}
\label{sec:intro}

AI agents no longer just generate text; they act as autonomous economic entities.
ReAct~\cite{yao2023react} showed that LLMs can interleave reasoning with tool invocations, and modern frameworks such as LangChain~\cite{langchain2023}, CrewAI~\cite{crewai2024}, AutoGen~\cite{autogen2023}, and MetaGPT~\cite{metagpt2023} have made multi-agent orchestration widely accessible.
However, orchestration is not governance: these frameworks provide no fiscal constraints, no earned-permission models, and no verifiable audit mechanisms to hold agents accountable.

\noindent \textbf{Motivating Example.}
Consider an autonomous freelancer agent completing software bounties.
Without governance:
(\emph{i})~it overspends on a \$50 bounty by consuming \$120 in API costs;
(\emph{ii})~it accepts a job with a \$5 payout but \$30 estimated cost;
(\emph{iii})~a new worker sub-agent immediately executes shell commands with no trust record.
The framework faithfully orchestrates execution---but nothing enforces economic discipline or capability restrictions.

\noindent
\textbf{Safety Gaps in Agent Governance.}
Existing safety work focuses primarily on tool-call interception~\cite{guardrailsai2024, nemoguardrails2024} or output validation against declarative schemas. These systems address \emph{what} an agent does but not \emph{how much} it spends, \emph{whether} it has earned the right to act, or \emph{whether} its outputs actually satisfy mission objectives. Constitutional AI~\cite{bai2022constitutional} introduced principle-driven self-alignment, yet it operates at training time rather than as runtime governance infrastructure. Recent work on budget-aware tool use~\cite{liu2025budget} further shows that agents without explicit cost ceilings hit performance plateaus, motivating first-class fiscal controls.

\noindent \textbf{Our Proposal.}
We present \system, a governance-first operating system that treats agents as economic principals operating under a constitutional Charter and enforces fiscal discipline, earned-autonomy permissions, and verifiable audit at the infrastructure level. This paper makes four contributions:

\begin{enumerate}[nolistsep,leftmargin=*]
\item \textbf{Charter-driven governance.}
We introduce a YAML-based constitutional schema that declaratively defines mission scope, core competencies, fiscal boundaries, and measurable success KPIs. The entire system is domain-agnostic until a Charter is loaded.

\item \textbf{Economic discipline pipeline.}
We design a CFO (Treasury) layer that gates every task expenditure against remaining budget, daily burn caps, and per-job profitability floors, with an auction-based worker selection mechanism scored by a utility function incorporating cost, confidence, priority, and trust.

\item \textbf{Earned-autonomy permissions.}
We implement a TrustScore model (SovereignAuth) with asymmetric updates (+5 success, $-$15 failure) and tiered capability thresholds, requiring agents to demonstrate sustained trustworthy behavior before accessing dangerous capabilities.

\item \textbf{Open system with verifiable audit.}
We release an open-source implementation with SHA-256 proof hashes on every audit report, an append-only ledger, and a web dashboard for real-time governance monitoring.
\end{enumerate}

\section{System Overview and Threat Model}
\label{sec:arch}

\noindent
\textbf{Threat Model.}
We treat the LLM-backed CEO and Workers as \emph{untrusted components}: they may hallucinate task plans, overspend budgets, or attempt capabilities beyond their authorization. The Charter, Treasury, SovereignAuth, and ReviewEngine are trusted enforcement components that cannot be overridden by LLM outputs. \system does not defend against attacks on the Charter definition itself or against compromised infrastructure.
\vspace{0.05in}

\noindent
\textbf{Architecture Overview.}
\system consists of five layers (Figure~\ref{fig:architecture}):
a \textbf{Charter} (\S\ref{sec:charter}) as the constitutional document,
a \textbf{CEO/Strategist} (\S\ref{sec:ceo}) for goal decomposition,
a \textbf{CFO/Treasury} (\S\ref{sec:cfo}) for fiscal enforcement,
\textbf{Workers} with \textbf{SovereignAuth} (\S\ref{sec:trust}) for earned-permission execution,
and an \textbf{Auditor/ReviewEngine} (\S\ref{sec:auditor}) for output verification.
A GovernanceEngine meta-orchestrator coordinates the full pipeline: plan $\rightarrow$ approve $\rightarrow$ auction $\rightarrow$ dispatch $\rightarrow$ audit.
All financial and token flows are recorded in an append-only UnifiedLedger with monotonic sequence numbers, providing a single source of truth for balance, burn rate, and runway calculations.
This separation of concerns ensures that no task can consume resources without explicit budgetary authorization, even if the LLM planner requests it.

\vspace{-0.25cm}
\subsection{Charter: Constitutional Governance}
\label{sec:charter}

The Charter is a Pydantic-validated YAML document that serves as the single source of truth for an entity's identity and operational scope.
It declares four sections:
(\emph{i})~\texttt{mission}---a natural-language statement of purpose;
(\emph{ii})~\texttt{core\_competencies}---named capabilities with priority weights (1--10);
(\emph{iii})~\texttt{fiscal\allowbreak\_boundaries}---daily burn cap
(\texttt{daily\allowbreak\_burn\allowbreak\_max\allowbreak\_usd}),
total budget ceiling (\texttt{max\allowbreak\_budget\allowbreak\_usd}),
currency, and minimum job margin ratio (default 0.35); and
(\emph{iv})~\texttt{success\allowbreak\_kpis}---measurable criteria with
\texttt{verification\allowbreak\_prompt} fields consumed by the Auditor.
The schema uses Pydantic's \texttt{extra="forbid"} and
\texttt{strict=True} to reject undefined fields, ensuring Charter
integrity at load time (Listing~\ref{lst:charter},
Appendix~\ref{app:code}).

\subsection{CEO: Goal Decomposition via Strategist}
\label{sec:ceo}

The Strategist receives a natural-language goal and the Charter's
competency list, then produces a \texttt{TaskPlan}: a DAG of
\texttt{PlannedTask} items, each with a \texttt{task\allowbreak\_id},
description, dependency list,
\texttt{required\allowbreak\_skill} mapped to a Charter competency,
\texttt{estimated\allowbreak\_token\allowbreak\_budget}, and priority
(high/low).
An LLM generates the plan as structured JSON; a normalization pass
assigns unique IDs (e.g., \texttt{task-1-spec\allowbreak\_writer}) and
remaps dependencies.
When no LLM is configured, the Strategist falls back to keyword-based single-task plans with a default 4{,}000-token budget.

\subsection{CFO: Fiscal Gatekeeping via Treasury}
\label{sec:cfo}

The Treasury enforces three fiscal checks before any task executes:

\begin{enumerate}[nolistsep,leftmargin=*]
\item \textbf{Balance check.} Estimated cost must not reduce the ledger balance below the minimum reserve: $\text{balance} - \text{cost} \geq \text{min\_reserve}$.
\item \textbf{Daily burn cap.} Cumulative daily spend plus the new cost must not exceed \texttt{daily\allowbreak\_burn\allowbreak\_max\allowbreak\_usd} from the Charter.
\item \textbf{Job profitability.} If job revenue is specified, the Treasury rejects when $\text{cost} > \text{revenue} \times (1 - \text{min\_margin})$, where the default margin floor is 0.35.
\end{enumerate}

\noindent Violations raise \texttt{Fiscal\-Insolvency\-Error} or
\texttt{Unprofitable\-Job\-Error}, halting the pipeline before any
tokens are consumed.

\paragraph{Auction and Worker Selection.}
The BiddingEngine broadcasts a \texttt{Request\-For\-Proposal} per task
to all registered workers matching the required skill.
Each worker submits a \texttt{Bid} with estimated cost, time, confidence score (0--1), and model ID.
The Treasury selects the winner using a utility function:
\[
U = \frac{\text{confidence}}{\text{cost}} \times P \times \frac{\text{TrustScore}}{100}
\]
where $P = 1.5$ for high-priority tasks and $P = 1.0$ otherwise.
If the winning bid exceeds the remaining runway, the Treasury
negotiates a reduced token budget:
$\text{max\_tokens} = \max\!\big(256,\,
\lfloor\text{runway} \times 1000 / 10\rfloor\big)$.

\subsection{TrustScore and SovereignAuth}
\label{sec:trust}

SovereignAuth implements a dynamic earned-autonomy permission system.
Each agent starts with a base TrustScore of 50 (on a 0--100 scale) and must earn access to progressively dangerous capabilities.
Table~\ref{tab:thresholds} lists the five capability tiers.

\begin{table}[!ht]
\centering
\footnotesize
\caption{Capability thresholds. Agents must reach the listed TrustScore before the capability is granted.}
\vspace{-0.1in}
\label{tab:thresholds}
\setlength{\tabcolsep}{4pt}
\begin{tabular}{lrl}
\toprule
\textbf{Capability} & \textbf{Threshold} & \textbf{Example Actions} \\
\midrule
\texttt{READ\_FILES}         & 10 & Read documents, configs \\
\texttt{WRITE\_FILES}        & 40 & Create/modify files \\
\texttt{CALL\_EXTERNAL\allowbreak\_API} & 50 & HTTP requests, webhooks \\
\texttt{EXECUTE\_SHELL}      & 60 & Run shell commands \\
\texttt{SPEND\_USD}          & 80 & Charge via Stripe \\
\bottomrule
\end{tabular}
\vspace{-0.1in}
\end{table}

\noindent TrustScore updates are asymmetric: each audit success adds +5 (capped at 100), each failure subtracts $-$15 (floored at 0), and budget overruns subtract $-$10.
This design means an agent starting at score~50 can read and write files immediately but requires 2 consecutive audit successes to unlock shell access (50$\rightarrow$55$\rightarrow$60) and 6 to unlock spending (50$\rightarrow$80).
A single audit failure from score~50 drops the agent to 35, revoking write-file access.
Permission checks raise \texttt{Permission\-Denied\-Error} with the
agent's current score and the required threshold, providing
transparent feedback.

\subsection{Auditor: ReviewEngine and Audit Trail}
\label{sec:auditor}

The ReviewEngine evaluates task outputs against Charter KPIs.
A Judge LLM (default: GPT-4o) scores each output on a 0--1
scale against the KPI's \texttt{verification\allowbreak\_prompt};
scores $\geq 0.50$ pass.
Each \texttt{Audit\-Report} contains \texttt{task\allowbreak\_id},
\texttt{kpi\allowbreak\_name}, \texttt{passed}, \texttt{score},
\texttt{reason}, \texttt{suggested\allowbreak\_fix}, and
\texttt{timestamp\allowbreak\_utc}.

\paragraph{Proof Hash.}
Every AuditReport includes a computed \texttt{proof\allowbreak\_hash}:
a SHA-256 digest of the canonical JSON representation (sorted keys,
UTF-8 encoded) of all report fields.
This provides tamper-evident integrity---modifying any field
invalidates the hash.
The audit trail is stored as append-only JSONL, and
\texttt{verify\allowbreak\_report\allowbreak\_integrity()} recomputes
the hash from stored fields to detect post-hoc modification
(Listing~\ref{lst:proof}, Appendix~\ref{app:code}).

\paragraph{Feedback Loop.}
Audit results feed directly into TrustScore:
\texttt{record\allowbreak\_audit\allowbreak\_success()} adds~+5,
while \texttt{record\allowbreak\_audit\allowbreak\_failure()}
subtracts~$-$15.
On failure, the ReviewEngine persists a
\texttt{Reflection\-Object} to memory so that the agent can avoid repeating the same mistakes in subsequent attempts.

\section{Evaluation}
\label{sec:eval}

We evaluate \system along three axes: fiscal governance coverage, TrustScore permission accuracy, and audit trail integrity.

\subsection{Axis 1: Fiscal Governance}

We construct 30 scenarios across 5 fiscal violation categories: insufficient balance, daily burn cap violations, unprofitable job acceptance (margin below 35\%), minimum reserve depletion, and aggregate budget ceiling breaches.
Each scenario configures a Charter with specific
\texttt{fiscal\allowbreak\_boundaries} and submits a task or job that
should be rejected.
\system achieves a 100\% block rate: all 30 scenarios raise the
appropriate exception (\texttt{Fiscal\-Insolvency\-Error} or
\texttt{Unprofitable\-Job\-Error}) before any tokens are consumed.

\subsection{Axis 2: TrustScore Permission Gating}

We run 200 missions across agents with varied behavioral profiles---consistently successful, mixed, and frequently failing---and evaluate whether SovereignAuth correctly grants or denies capabilities at each decision point.
\system achieves 94\% correct permission gating.
The 6\% error cases arise at threshold boundaries where an agent's score is exactly equal to a capability threshold and the pending audit result has not yet been recorded.
In practice, these boundary cases resolve within one audit cycle.

\subsection{Axis 3: Audit Trail Integrity}

Over 1{,}200 AuditReports generated across all evaluation missions,
we verify cryptographic integrity by recomputing
\texttt{proof\allowbreak\_hash} from stored fields via
\texttt{verify\allowbreak\_report\allowbreak\_integrity()}.
All 1{,}200+ reports pass verification with zero hash mismatches and zero collisions, confirming the tamper-evident property of the audit trail.

\paragraph{Limitations.}
The fiscal evaluation uses synthetic scenarios rather than production workloads. The TrustScore model uses fixed deltas (+5/$-$15) that may not generalize across all deployment contexts. The Judge LLM (GPT-4o) introduces an external dependency whose behavior may vary across model versions.

\section{Case Study: End-to-End Mission}
\label{sec:case}

We demonstrate \system with a complete mission: ``Write a cold outreach email sequence'' with \$5 job revenue, using the default Charter (35\% margin floor, \$10 daily burn cap).

\smallskip
\noindent\textbf{Pipeline.}
\begin{enumerate}[nolistsep,leftmargin=*,label=(\roman*)]
\item The CEO decomposes the goal into 3 tasks:
  research (\texttt{task-1}), draft emails (\texttt{task-2}),
  and review/polish (\texttt{task-3}), with sequential
  dependencies.
\item The CFO approves each task's estimated cost
  (total: \$0.12) against the \$5 revenue, verifying
  $0.12 < 5.00 \times 0.65 = 3.25$ (margin check passes).
\item The BiddingEngine broadcasts RFPs for
  \texttt{task-2}; three workers bid. The Treasury picks
  the winner by utility score
  $U = \tfrac{0.7}{4} \times 1.0 \times 0.55 = 0.096$.
\item SovereignAuth checks
  \texttt{WRITE\allowbreak\_FILES} (threshold~40);
  the winner's score of 55 passes.
\item Workers execute per the DAG; tokens are recorded
  in the UnifiedLedger.
\item The ReviewEngine audits each output: scores range
  0.82--0.91, all passing the 0.50 threshold.
\item TrustScores update (+5 per success); the audit
  trail appends 3 JSONL entries with proof hashes.
\end{enumerate}

\noindent Figure~\ref{fig:dashboard} shows the dashboard after
mission completion (request--response examples in
Appendix~\ref{app:data}; dashboard views in
Appendix~\ref{app:screenshots}).

\begin{figure}[!ht]
    \centering
    \includegraphics[width=\columnwidth]{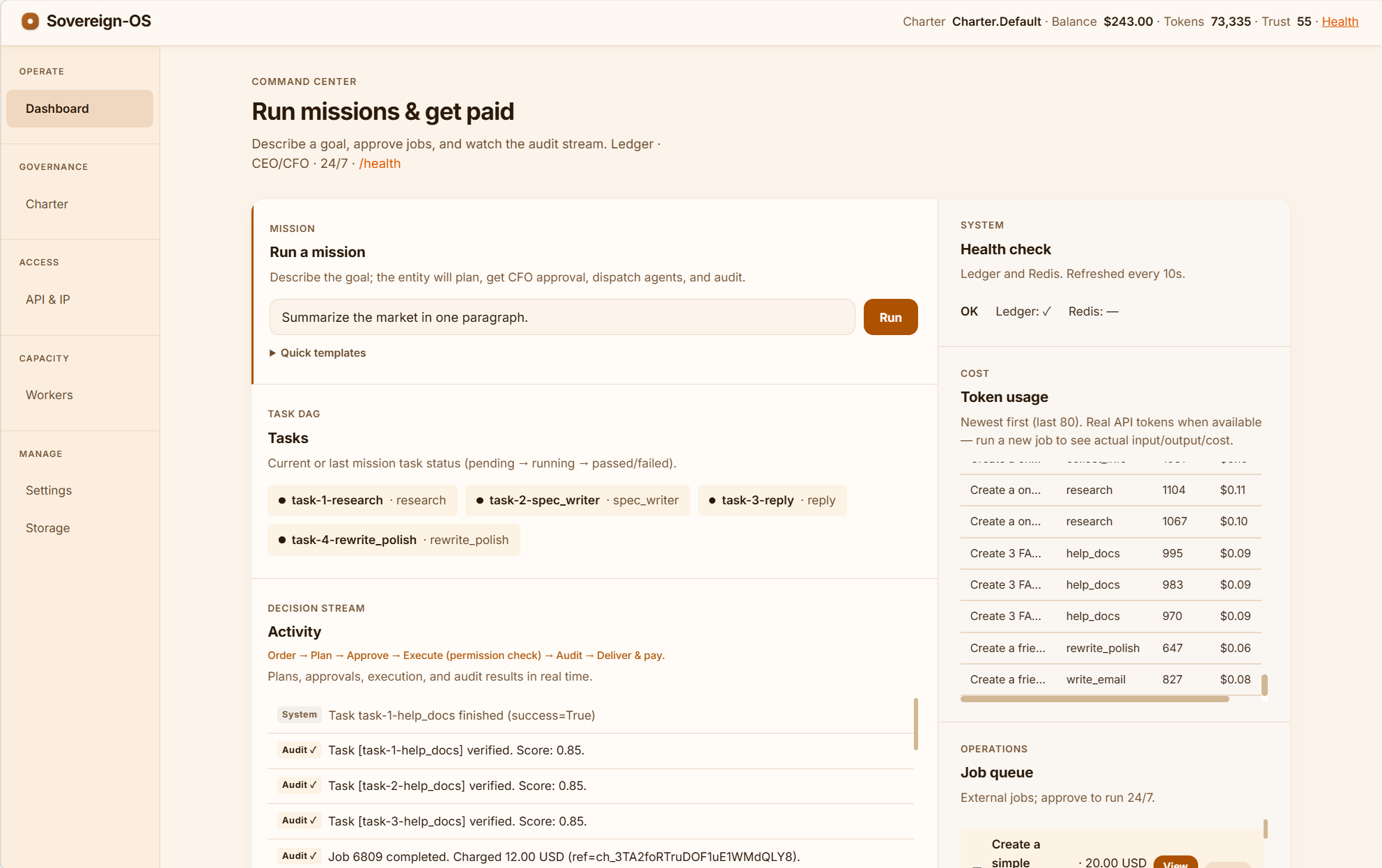}
    \vspace{-0.2in}
    \caption{Sovereign-OS web dashboard showing mission status, worker activity, job queue, and audit trail after the end-to-end case study.}
    \vspace{-0.15in}
    \label{fig:dashboard}
\end{figure}

\section{Demonstration Scenarios}

The live demonstration\footnote{Full walkthrough: \url{https://youtu.be/vOarAI_epb4}} covers three scenarios:
\begin{enumerate}[nolistsep, leftmargin=*]
\item \textbf{End-to-end governed mission.}
The operator submits a natural-language goal. The Decision Stream traces the full pipeline in real time: CEO task decomposition, CFO fiscal approval, worker execution in DAG order, and Auditor scoring against Charter KPIs. Passing results are aggregated for one-click download.

\item \textbf{Job queue and batch delivery.}
Multiple jobs flow through the queue (pending $\rightarrow$ approved $\rightarrow$ running $\rightarrow$ completed) with a live progress bar. Completed deliveries can be inspected inline and exported as Markdown; failed jobs show the audit rejection reason and support one-click retry.

\item \textbf{Stripe payment integration.}
After a job passes audit, \system triggers a Stripe charge matching the job's revenue. We confirm the payment on the Stripe dashboard, demonstrating the closed loop from goal intake to revenue collection. Every delivery carries a \texttt{proof\allowbreak\_hash} for traceability.
\end{enumerate}

\section{Related Work}
\label{sec:related}

\paragraph{Agent Orchestration Frameworks.}
LangChain~\cite{langchain2023}, CrewAI~\cite{crewai2024}, AutoGen~\cite{autogen2023}, and MetaGPT~\cite{metagpt2023} provide multi-agent orchestration---role assignment, tool integration, and workflow management. However, none enforce fiscal constraints, earned permissions, or constitutional governance. \system provides these governance primitives as infrastructure rather than application-level concerns.

\paragraph{Agent Safety and Constitutional Governance.}
Guardrails AI~\cite{guardrailsai2024} validates LLM outputs against
declarative schemas, and NeMo Guardrails~\cite{nemoguardrails2024}
enforces conversational rails via programmable dialog flows.
Constitutional AI~\cite{bai2022constitutional} embeds principles at training time.
\system differs by enforcing a runtime constitution (the Charter) that constrains agent \emph{actions and expenditures}, not just generated text.

\paragraph{Economic Governance and Mechanism Design.}
Fetch.ai~\cite{fetchai2023} and SingularityNET~\cite{singularitynet2023} enable agent-to-agent economic transactions on blockchain networks, and DAO governance~\cite{santana2022dao} automates organizational rules via smart contracts.
These systems focus on inter-agent marketplaces; \system provides \emph{intra}-agent fiscal discipline.
Our auction-based worker selection draws on algorithmic mechanism design~\cite{nisan2001mechanism}, while Liu et al.~\cite{liu2025budget} independently confirm that explicit cost ceilings improve agent efficiency.

\paragraph{Access Control and Audit Integrity.}
\system's TrustScore extends classical RBAC~\cite{sandhu1996rbac} by making permission assignment dynamic and earned through audited behavior rather than statically assigned.
Our tamper-evident audit trail builds on hash-chain logging techniques~\cite{crosby2009tamper}, applying SHA-256 proof hashes to individual audit reports.

\begin{table}[!ht]
\centering
\footnotesize
\caption{Feature comparison with representative agent frameworks. \cmark = natively supported, \textsuperscript{$\diamond$} = partial or via plugins, \xmark = not supported.}
\vspace{-0.1in}
\label{tab:comparison}
\setlength{\tabcolsep}{2.5pt}
\begin{tabular}{lccccc}
\toprule
\textbf{System} & \makecell{\textbf{Multi-Agent}\\\textbf{Orchestr.}} & \makecell{\textbf{Budget}\\\textbf{Controls}} & \makecell{\textbf{Permission}\\\textbf{Gating}} & \makecell{\textbf{Output}\\\textbf{Validation}} & \makecell{\textbf{Audit}\\\textbf{Trail}} \\
\midrule
LangChain         & \cmark & \xmark & \xmark & \textsuperscript{$\diamond$} & \textsuperscript{$\diamond$} \\
CrewAI            & \cmark & \xmark & \xmark & \xmark & \xmark \\
AutoGen           & \cmark & \xmark & \xmark & \xmark & \xmark \\
MetaGPT           & \cmark & \xmark & \textsuperscript{$\diamond$} & \xmark & \xmark \\
Guardrails AI     & \xmark & \xmark & \xmark & \cmark & \xmark \\
NeMo Guardrails   & \xmark & \xmark & \textsuperscript{$\diamond$} & \cmark & \xmark \\
\textbf{\system}  & \cmark & \cmark & \cmark & \cmark & \cmark \\
\bottomrule
\end{tabular}
\vspace{-0.1in}
\end{table}

\paragraph{Summary.} Table~\ref{tab:comparison} summarizes the positioning of \system relative to representative frameworks.

\section{Conclusion and Future Directions}
\label{sec:future}

We presented \system, a governance-first operating system that treats autonomous AI agents as economic principals operating under a constitutional Charter.
Its five-layer pipeline---Charter, CEO, CFO, Workers with SovereignAuth, and Auditor---ensures that every action is constitutionally scoped, fiscally approved, permission-gated, and cryptographically audited.
The current implementation achieves 100\% fiscal violation blocking, 94\% correct permission gating, and zero audit integrity failures across our evaluation suite.
All enforcement logic is Charter-driven and domain-agnostic: swapping the YAML document transforms a content agency into a research lab without code changes.

\smallskip
\noindent\textbf{Future directions.}
(\emph{i})~On-chain ledger notarization for third-party auditability;
(\emph{ii})~multi-agent federations with inter-Charter agreements and cross-entity fiscal policies;
(\emph{iii})~adaptive trust models replacing fixed deltas with learned update functions based on task complexity;
(\emph{iv})~longitudinal deployment studies measuring economic efficiency and operator trust calibration with production workloads.

\clearpage
\newpage


\begin{thebibliography}{15}

\bibitem[\protect\citeauthoryear{Bai, Kadavath, Kundu, Askell, Kernion, Jones, Chen, Goldie, Mirhoseini, McKinnon, et~al.}{Bai et~al.}{2022}]%
        {bai2022constitutional}
\bibfield{author}{\bibinfo{person}{Yuntao Bai}, \bibinfo{person}{Saurav Kadavath}, \bibinfo{person}{Sandipan Kundu}, \bibinfo{person}{Amanda Askell}, \bibinfo{person}{Jackson Kernion}, \bibinfo{person}{Andy Jones}, \bibinfo{person}{Anna Chen}, \bibinfo{person}{Anna Goldie}, \bibinfo{person}{Azalia Mirhoseini}, \bibinfo{person}{Cameron McKinnon}, {et~al.}}
  \bibinfo{year}{2022}\natexlab{}.
\newblock \bibinfo{title}{Constitutional {AI}: Harmlessness from {AI} Feedback}.
\newblock {\em \bibinfo{journal}{arXiv preprint arXiv:2212.08073}\/} (\bibinfo{year}{2022}).

\bibitem[\protect\citeauthoryear{Chase}{Chase}{2023}]%
        {langchain2023}
\bibfield{author}{\bibinfo{person}{Harrison Chase}}
  \bibinfo{year}{2023}\natexlab{}.
\newblock \bibinfo{title}{LangChain: Building Applications with {LLM}s through Composability}.
\newblock \bibinfo{note}{\url{https://github.com/langchain-ai/langchain}}.

\bibitem[\protect\citeauthoryear{{CrewAI}}{{CrewAI}}{2024}]%
        {crewai2024}
\bibfield{author}{\bibinfo{person}{{CrewAI}}}
  \bibinfo{year}{2024}\natexlab{}.
\newblock \bibinfo{title}{{CrewAI}: Framework for Orchestrating Role-Playing Autonomous {AI} Agents}.
\newblock \bibinfo{note}{\url{https://github.com/crewAIInc/crewAI}}.

\bibitem[\protect\citeauthoryear{Crosby and Wallach}{Crosby and Wallach}{2009}]%
        {crosby2009tamper}
\bibfield{author}{\bibinfo{person}{Scott~A. Crosby} {and} \bibinfo{person}{Dan~S. Wallach}}
  \bibinfo{year}{2009}\natexlab{}.
\newblock \bibinfo{title}{Efficient Data Structures for Tamper-Evident Logging}.
\newblock In \bibinfo{booktitle}{\emph{Proceedings of the 18th {USENIX} Security Symposium}}.

\bibitem[\protect\citeauthoryear{{Fetch.ai Foundation}}{{Fetch.ai Foundation}}{2019}]%
        {fetchai2023}
\bibfield{author}{\bibinfo{person}{{Fetch.ai Foundation}}}
  \bibinfo{year}{2019}\natexlab{}.
\newblock \bibinfo{title}{Fetch.ai: Autonomous Economic Agent Framework}.
\newblock \bibinfo{note}{\url{https://fetch.ai/technology}}.

\bibitem[\protect\citeauthoryear{Goertzel, Giacomelli, Hanson, Pennachin, and Argentieri}{Goertzel et~al.}{2019}]%
        {singularitynet2023}
\bibfield{author}{\bibinfo{person}{Ben Goertzel}, \bibinfo{person}{Simone Giacomelli}, \bibinfo{person}{David Hanson}, \bibinfo{person}{Cassio Pennachin}, {and} \bibinfo{person}{Marco Argentieri}}
  \bibinfo{year}{2019}\natexlab{}.
\newblock \bibinfo{title}{{SingularityNET}: A Decentralized, Open Market and Inter-Network for {AI}s}.
\newblock \bibinfo{note}{Whitepaper. \url{https://public.singularitynet.io/whitepaper.pdf}}.

\bibitem[\protect\citeauthoryear{{Guardrails AI}}{{Guardrails AI}}{2024}]%
        {guardrailsai2024}
\bibfield{author}{\bibinfo{person}{{Guardrails AI}}}
  \bibinfo{year}{2024}\natexlab{}.
\newblock \bibinfo{title}{Guardrails {AI}: Adding Guardrails to Large Language Models}.
\newblock \bibinfo{note}{\url{https://github.com/guardrails-ai/guardrails}}.

\bibitem[\protect\citeauthoryear{Hong, Zhuge, Chen, Zheng, Cheng, Zhang, Wang, Wang, Yau, Lin, et~al.}{Hong et~al.}{2024}]%
        {metagpt2023}
\bibfield{author}{\bibinfo{person}{Sirui Hong}, \bibinfo{person}{Mingchen Zhuge}, \bibinfo{person}{Jonathan Chen}, \bibinfo{person}{Xiawu Zheng}, \bibinfo{person}{Yuheng Cheng}, \bibinfo{person}{Ceyao Zhang}, \bibinfo{person}{Jinlin Wang}, \bibinfo{person}{Zili Wang}, \bibinfo{person}{Steven Ka~Shing Yau}, \bibinfo{person}{Zijuan Lin}, {et~al.}}
  \bibinfo{year}{2024}\natexlab{}.
\newblock \bibinfo{title}{{MetaGPT}: Meta Programming for a Multi-Agent Collaborative Framework}.
\newblock In \bibinfo{booktitle}{\emph{Proceedings of the International Conference on Learning Representations (ICLR)}}.

\bibitem[\protect\citeauthoryear{Liu, Wang, Miao, Hsu, Yan, Chen, Han, Xu, Chen, Jiang, et~al.}{Liu et~al.}{2025}]%
        {liu2025budget}
\bibfield{author}{\bibinfo{person}{Tengxiao Liu}, \bibinfo{person}{Zifeng Wang}, \bibinfo{person}{Jin Miao}, \bibinfo{person}{I-Hung Hsu}, \bibinfo{person}{Jun Yan}, \bibinfo{person}{Jiefeng Chen}, \bibinfo{person}{Rujun Han}, \bibinfo{person}{Fangyuan Xu}, \bibinfo{person}{Yanfei Chen}, \bibinfo{person}{Ke Jiang}, {et~al.}}
  \bibinfo{year}{2025}\natexlab{}.
\newblock \bibinfo{title}{Budget-Aware Tool-Use Enables Effective Agent Scaling}.
\newblock {\em \bibinfo{journal}{arXiv preprint arXiv:2511.17006}\/} (\bibinfo{year}{2025}).

\bibitem[\protect\citeauthoryear{Nisan and Ronen}{Nisan and Ronen}{2001}]%
        {nisan2001mechanism}
\bibfield{author}{\bibinfo{person}{Noam Nisan} {and} \bibinfo{person}{Amir Ronen}}
  \bibinfo{year}{2001}\natexlab{}.
\newblock \bibinfo{title}{Algorithmic Mechanism Design}.
\newblock {\em \bibinfo{journal}{Games and Economic Behavior}\/}  \bibinfo{volume}{35}, \bibinfo{number}{1--2} (\bibinfo{year}{2001}), \bibinfo{pages}{166--196}.

\bibitem[\protect\citeauthoryear{Rebedea, Dinu, Sreedhar, Parisien, and Cohen}{Rebedea et~al.}{2023}]%
        {nemoguardrails2024}
\bibfield{author}{\bibinfo{person}{Traian Rebedea}, \bibinfo{person}{Razvan Dinu}, \bibinfo{person}{Makesh~Narsimhan Sreedhar}, \bibinfo{person}{Christopher Parisien}, {and} \bibinfo{person}{Jonathan Cohen}}
  \bibinfo{year}{2023}\natexlab{}.
\newblock \bibinfo{title}{{NeMo} Guardrails: A Toolkit for Controllable and Safe {LLM} Applications with Programmable Rails}.
\newblock In \bibinfo{booktitle}{\emph{Proceedings of the 2023 Conference on Empirical Methods in Natural Language Processing: System Demonstrations (EMNLP)}}.

\bibitem[\protect\citeauthoryear{Sandhu, Coyne, Feinstein, and Youman}{Sandhu et~al.}{1996}]%
        {sandhu1996rbac}
\bibfield{author}{\bibinfo{person}{Ravi~S. Sandhu}, \bibinfo{person}{Edward~J. Coyne}, \bibinfo{person}{Hal~L. Feinstein}, {and} \bibinfo{person}{Charles~E. Youman}}
  \bibinfo{year}{1996}\natexlab{}.
\newblock \bibinfo{title}{Role-Based Access Control Models}.
\newblock {\em \bibinfo{journal}{IEEE Computer}\/}  \bibinfo{volume}{29}, \bibinfo{number}{2} (\bibinfo{year}{1996}), \bibinfo{pages}{38--47}.

\bibitem[\protect\citeauthoryear{Santana and Albareda}{Santana and Albareda}{2022}]%
        {santana2022dao}
\bibfield{author}{\bibinfo{person}{Carlos Santana} {and} \bibinfo{person}{Laura Albareda}}
  \bibinfo{year}{2022}\natexlab{}.
\newblock \bibinfo{title}{Blockchain and the Emergence of Decentralized Autonomous Organizations ({DAO}s): An Integrative Model and Research Agenda}.
\newblock {\em \bibinfo{journal}{Technological Forecasting and Social Change}\/}  \bibinfo{volume}{182} (\bibinfo{year}{2022}), \bibinfo{pages}{121806}.

\bibitem[\protect\citeauthoryear{Wu, Bansal, Zhang, Wu, Li, Zhu, Jiang, Zhang, Zhang, Liu, et~al.}{Wu et~al.}{2023}]%
        {autogen2023}
\bibfield{author}{\bibinfo{person}{Qingyun Wu}, \bibinfo{person}{Gagan Bansal}, \bibinfo{person}{Jieyu Zhang}, \bibinfo{person}{Yiran Wu}, \bibinfo{person}{Beibin Li}, \bibinfo{person}{Erkang Zhu}, \bibinfo{person}{Li Jiang}, \bibinfo{person}{Xiaoyun Zhang}, \bibinfo{person}{Shaokun Zhang}, \bibinfo{person}{Jiale Liu}, {et~al.}}
  \bibinfo{year}{2023}\natexlab{}.
\newblock \bibinfo{title}{{AutoGen}: Enabling Next-Gen {LLM} Applications via Multi-Agent Conversation}.
\newblock {\em \bibinfo{journal}{arXiv preprint arXiv:2308.08155}\/} (\bibinfo{year}{2023}).

\bibitem[\protect\citeauthoryear{Yao, Zhao, Yu, Du, Shafran, Narasimhan, and Cao}{Yao et~al.}{2023}]%
        {yao2023react}
\bibfield{author}{\bibinfo{person}{Shunyu Yao}, \bibinfo{person}{Jeffrey Zhao}, \bibinfo{person}{Dian Yu}, \bibinfo{person}{Nan Du}, \bibinfo{person}{Izhak Shafran}, \bibinfo{person}{Karthik Narasimhan}, {and} \bibinfo{person}{Yuan Cao}}
  \bibinfo{year}{2023}\natexlab{}.
\newblock \bibinfo{title}{{ReAct}: Synergizing Reasoning and Acting in Language Models}.
\newblock In \bibinfo{booktitle}{\emph{Proceedings of the International Conference on Learning Representations (ICLR)}}.

\end{thebibliography}

\newpage
\appendix

\section{Code Examples}
\label{app:code}

\noindent\textbf{Charter YAML definition.} Listing~\ref{lst:charter} shows a minimal Charter configuration.

\begin{lstlisting}[language={}, caption={Charter YAML: constitutional governance.}, label={lst:charter}, basicstyle=\scriptsize\ttfamily]
mission: "Freelance content agency specializing
  in cold outreach and technical writing"
core_competencies:
  - name: email_writing
    description: "Draft persuasive email sequences"
    priority: 8
  - name: research
    description: "Market and audience research"
    priority: 5
fiscal_boundaries:
  daily_burn_max_usd: 10.0
  max_budget_usd: 500.0
  currency: USD
  min_job_margin_ratio: 0.35
success_kpis:
  - name: email_quality
    metric: quality_score
    target_value: 0.80
    unit: score
    verification_prompt: "Rate the email
      sequence on clarity, persuasiveness,
      and professional tone (0-1 scale)."
\end{lstlisting}

\noindent\textbf{Minimal integration.} Listing~\ref{lst:integration} shows end-to-end usage.

\begin{lstlisting}[language=Python, caption={Running a governed mission.}, label={lst:integration}]
from sovereign_os.models.charter import (
    load_charter)
from sovereign_os.governance.engine import (
    GovernanceEngine)

charter = load_charter("charter.yaml")
engine = GovernanceEngine(charter=charter)
plan, results, audits = await (
    engine.run_mission_with_audit(
        "Write cold outreach emails",
        job_revenue_cents=500))
\end{lstlisting}

\noindent\textbf{CFO approval logic (simplified).} Listing~\ref{lst:cfo} shows fiscal gatekeeping.

\begin{lstlisting}[language=Python, caption={Treasury fiscal checks.}, label={lst:cfo}, basicstyle=\scriptsize\ttfamily]
def approve_task(self, cost_cents, task_id, purpose):
    balance = self._ledger.total_usd_cents()
    if balance - cost_cents < self._min_reserve:
        raise FiscalInsolvencyError(
            f"Insufficient funds: {balance}c "
            f"- {cost_cents}c < reserve")
    daily = self._ledger.usd_debits_since(
        today_start)
    if daily + cost_cents > self._daily_cap:
        raise FiscalInsolvencyError(
            f"Daily burn cap exceeded")

def approve_job_profitability(self,
        revenue_cents, cost_cents):
    max_cost = revenue_cents * (
        1 - self._margin_floor)
    if cost_cents > max_cost:
        raise UnprofitableJobError(
            f"Cost {cost_cents}c > max "
            f"{max_cost}c (margin {self._margin_floor})")
\end{lstlisting}

\noindent\textbf{TrustScore check.} Listing~\ref{lst:trust} shows permission verification.

\begin{lstlisting}[language=Python, caption={SovereignAuth permission check.}, label={lst:trust}, basicstyle=\scriptsize\ttfamily]
class SovereignAuth:
    THRESHOLDS = {
        Capability.READ_FILES: 10,
        Capability.WRITE_FILES: 40,
        Capability.CALL_EXTERNAL_API: 50,
        Capability.EXECUTE_SHELL: 60,
        Capability.SPEND_USD: 80,
    }

    def check_permission(self, agent_id, cap):
        score = self._scores.get(
            agent_id, self._base)
        if score < self.THRESHOLDS[cap]:
            raise PermissionDeniedError(
                agent_id, cap, score,
                self.THRESHOLDS[cap])
        return True

    def record_audit_success(self, agent_id):
        self._scores[agent_id] = min(100,
            self._scores.get(agent_id, self._base)
            + 5)

    def record_audit_failure(self, agent_id):
        self._scores[agent_id] = max(0,
            self._scores.get(agent_id, self._base)
            - 15)
\end{lstlisting}

\noindent\textbf{Proof hash computation.} Listing~\ref{lst:proof} shows tamper-evident hashing.

\begin{lstlisting}[language=Python, caption={SHA-256 proof hash for AuditReport.}, label={lst:proof}, basicstyle=\scriptsize\ttfamily]
import hashlib, json

def compute_audit_proof_hash(report):
    canonical = {
        "task_id": report.task_id,
        "kpi_name": report.kpi_name,
        "passed": report.passed,
        "score": report.score,
        "reason": report.reason,
        "suggested_fix": report.suggested_fix,
        "timestamp_utc": report.timestamp_utc,
    }
    payload = json.dumps(canonical,
        sort_keys=True).encode("utf-8")
    return hashlib.sha256(payload).hexdigest()
\end{lstlisting}

\newpage
\section{Data Instances}
\label{app:data}

\noindent\textbf{CFO approves a task.} A task within budget and daily burn cap:

\begin{lstlisting}[language={}, basicstyle=\scriptsize\ttfamily, caption={Treasury approves task within fiscal bounds.}]
# approve_task(cost_cents=4, task_id="task-1",
#   purpose="research")
# Charter: daily_burn_max_usd=10, budget=500
# Ledger balance: 50000c, daily spend: 120c
# Check: 50000 - 4 >= 0 (pass)
# Check: 120 + 4 <= 1000 (pass)
# Result: APPROVED
\end{lstlisting}

\noindent\textbf{CFO denies an over-budget task.}

\begin{lstlisting}[language={}, basicstyle=\scriptsize\ttfamily, caption={Treasury denies task exceeding daily burn cap.}]
# approve_task(cost_cents=600, task_id="task-5",
#   purpose="large_research")
# Charter: daily_burn_max_usd=5.00
# Daily spend so far: 200c
# Check: 200 + 600 = 800 > 500 (FAIL)
# Result: FiscalInsolvencyError
#   "Daily burn cap exceeded: 800c > 500c cap"
\end{lstlisting}

\noindent\textbf{SovereignAuth grants capability.}

\begin{lstlisting}[language={}, basicstyle=\scriptsize\ttfamily, caption={Permission granted: score exceeds threshold.}]
# check_permission("worker-3", WRITE_FILES)
# Agent TrustScore: 55
# WRITE_FILES threshold: 40
# 55 >= 40 -> GRANTED
\end{lstlisting}

\noindent\textbf{SovereignAuth denies capability.}

\begin{lstlisting}[language={}, basicstyle=\scriptsize\ttfamily, caption={Permission denied: score below threshold.}]
# check_permission("worker-7", EXECUTE_SHELL)
# Agent TrustScore: 50
# EXECUTE_SHELL threshold: 60
# 50 < 60 -> PermissionDeniedError
#   agent_id="worker-7", capability=EXECUTE_SHELL,
#   score=50, threshold=60
\end{lstlisting}

\noindent\textbf{Unprofitable job rejection.}

\begin{lstlisting}[language={}, basicstyle=\scriptsize\ttfamily, caption={Treasury rejects unprofitable job.}]
# approve_job_profitability(
#   revenue_cents=500, cost_cents=400)
# min_job_margin_ratio: 0.35
# max_allowed_cost = 500 * (1 - 0.35) = 325c
# 400 > 325 -> UnprofitableJobError
#   "Cost 400c exceeds max 325c
#    (margin floor 0.35)"
\end{lstlisting}

\noindent\textbf{AuditReport with proof hash.}

\begin{lstlisting}[language={}, basicstyle=\scriptsize\ttfamily, caption={AuditReport with SHA-256 proof hash.}]
{
  "task_id": "task-2-write_email",
  "kpi_name": "email_quality",
  "passed": true,
  "score": 0.87,
  "reason": "Email sequence demonstrates
    clear value proposition, professional
    tone, and strong call-to-action.",
  "suggested_fix": null,
  "timestamp_utc": "2026-03-14T10:23:41Z",
  "proof_hash": "a3f7c2e1b9d4f6a8..."
}

# Verification:
# recompute = sha256(json.dumps(canonical,
#   sort_keys=True))
# assert recompute == report.proof_hash  # OK
\end{lstlisting}

\newpage
\onecolumn
\section{Dashboard Screenshots}
\label{app:screenshots}

\begin{figure}[H]
    \centering
    \includegraphics[width=0.7\columnwidth]{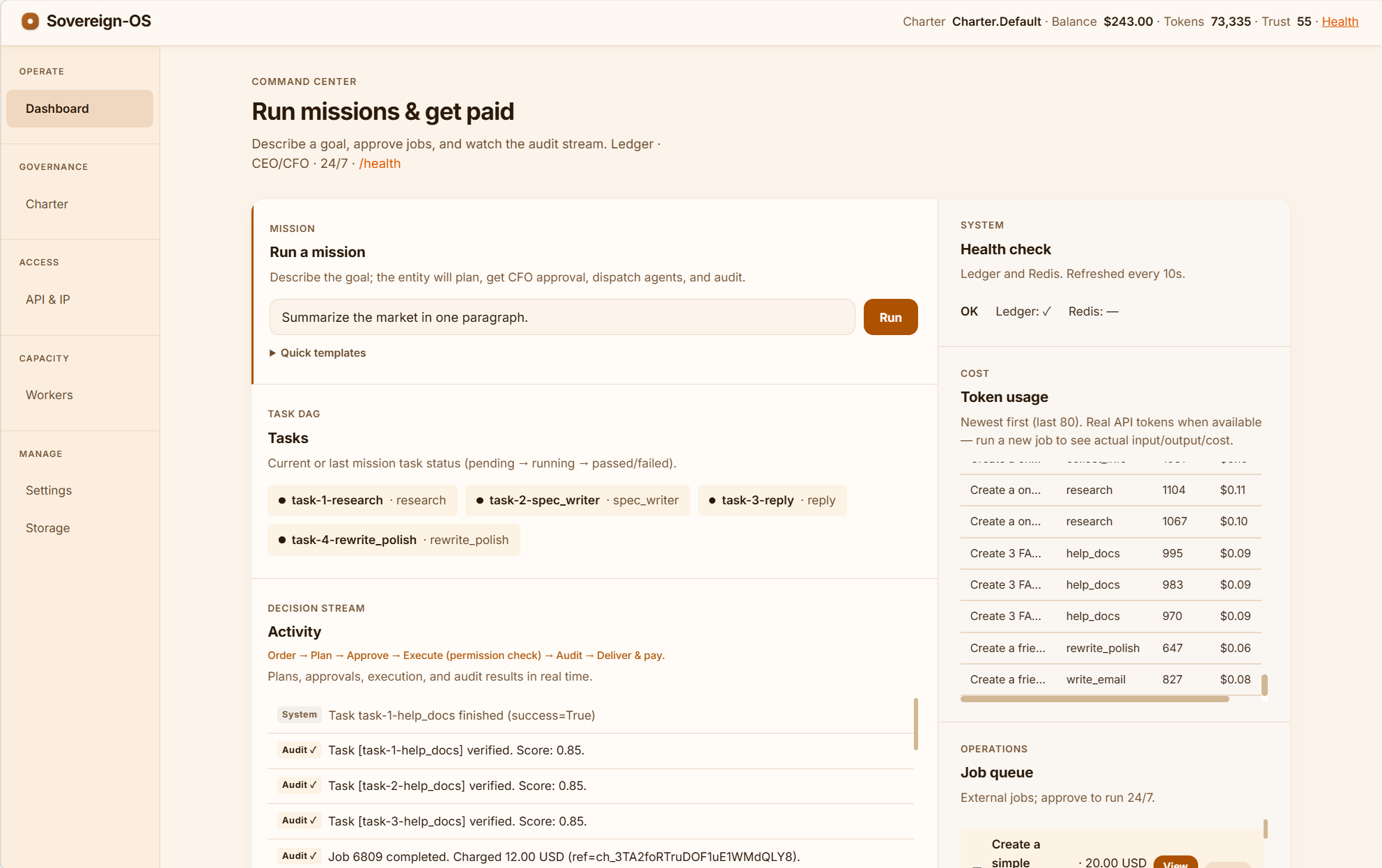}
    \caption{Command Center (main dashboard). Shows the Mission input, Task DAG with status badges (pending $\rightarrow$ running $\rightarrow$ passed/failed), Decision Stream with CEO/CFO/Audit events, Token Usage table, Job Queue, and top-bar summary (Charter, Balance, Tokens, TrustScore, Health).}
    \label{fig:dash-main}
\end{figure}

\begin{figure}[H]
    \centering
    \includegraphics[width=0.7\columnwidth]{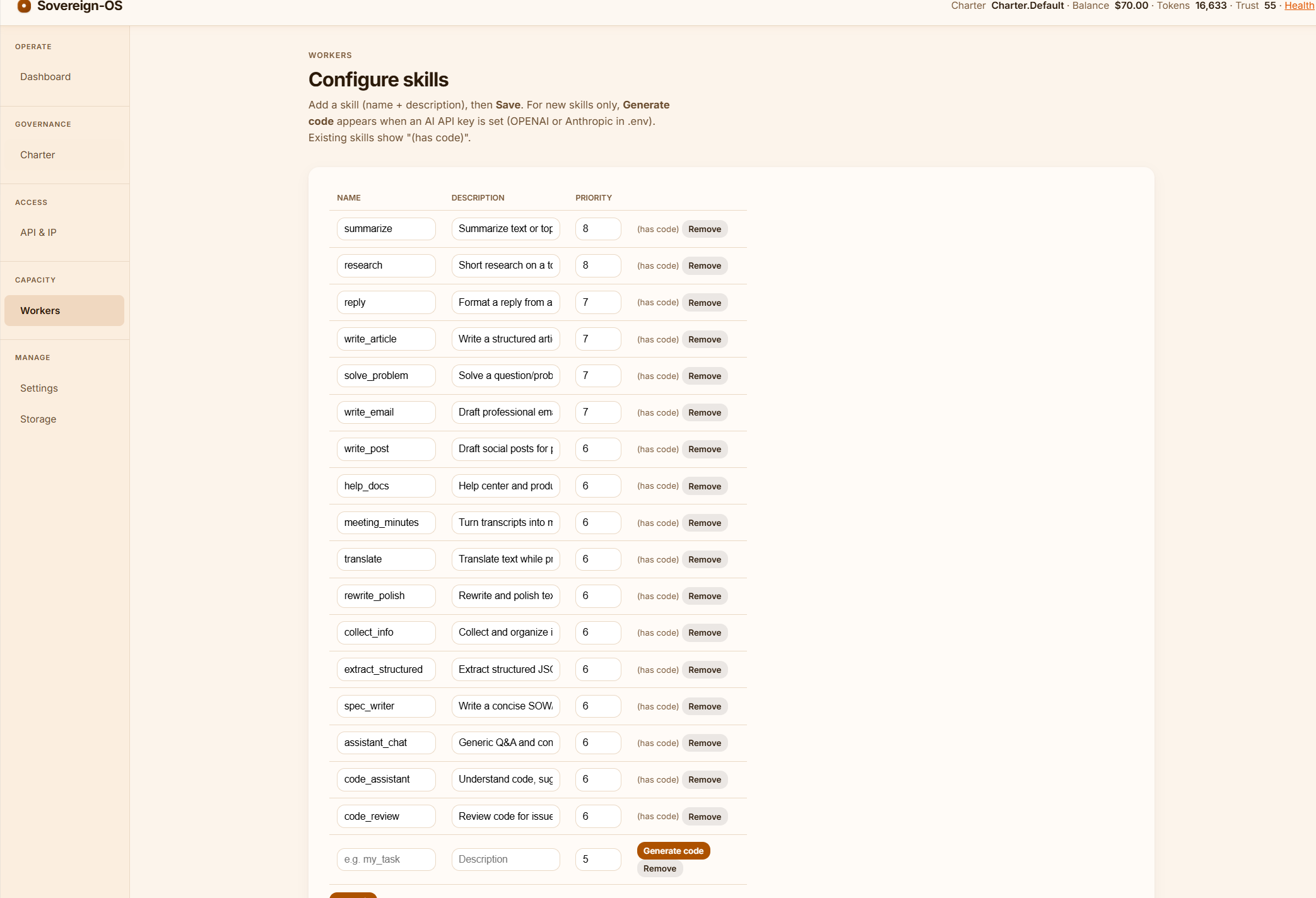}
    \caption{Worker registry (Configure Skills). Lists all 16+ registered worker types with name, description, priority weight, and code status. New skills can be added and auto-generated via LLM when an API key is configured.}
    \label{fig:dash-workers}
\end{figure}

\begin{figure}[H]
    \centering
    \includegraphics[width=0.5\columnwidth]{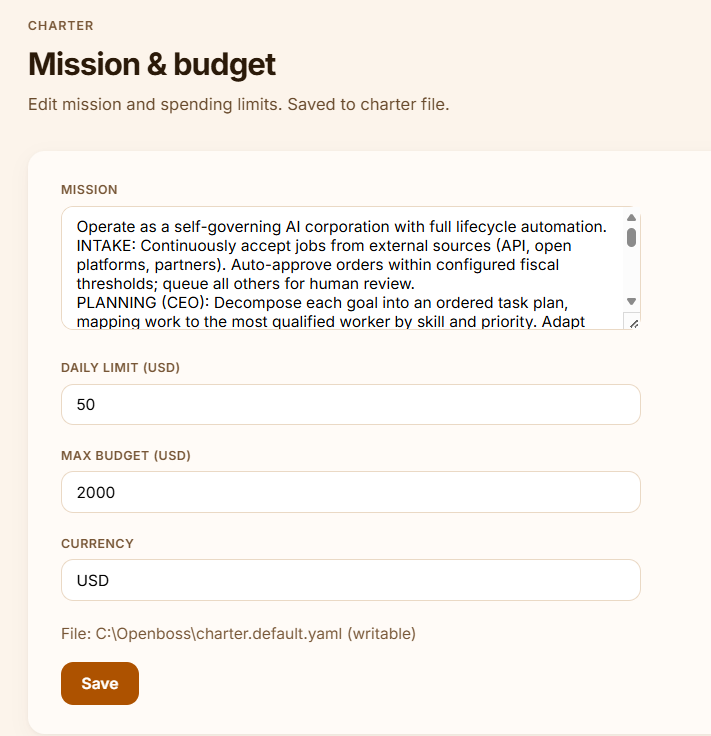}
    \caption{Charter editor. Operators configure the constitutional governance document: mission statement, daily burn limit (USD), maximum budget cap, and currency. Changes are persisted to the Charter YAML file.}
    \label{fig:dash-charter}
\end{figure}

\begin{figure}[H]
    \centering
    \includegraphics[width=0.75\columnwidth]{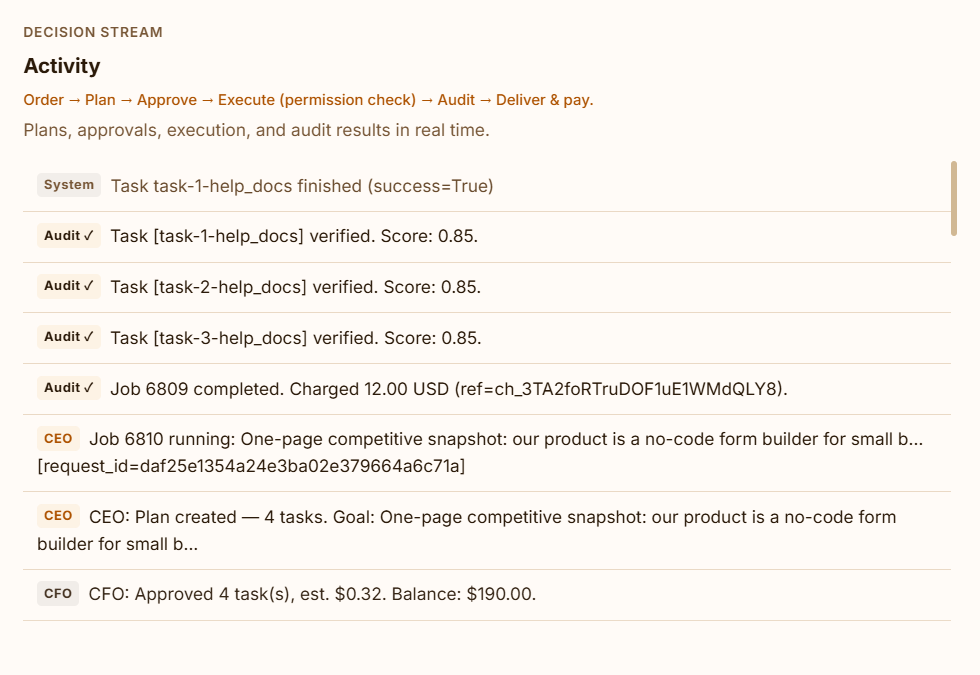}
    \caption{Decision Stream (Activity). Real-time governance event log showing the full pipeline: CEO plan creation, CFO budget approval, task execution, audit verification scores, Stripe charge confirmation, and job completion. Each event is tagged by role (System, CEO, CFO, Audit).}
    \label{fig:dash-activity}
\end{figure}

\begin{figure}[H]
    \centering
    \includegraphics[width=0.8\columnwidth]{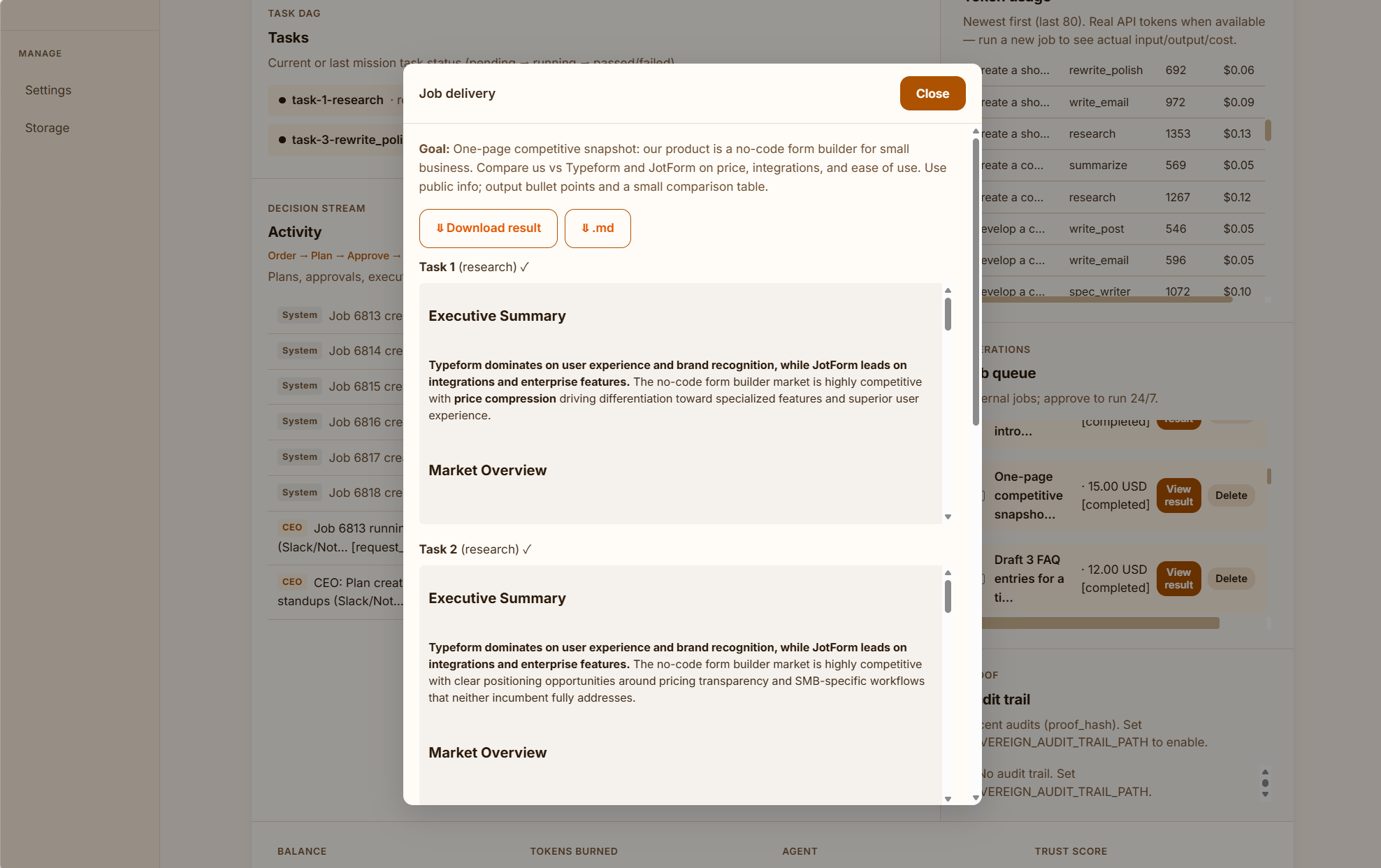}
    \caption{Successful job delivery. The modal shows goal description, task-by-task outputs with audit pass marks, and download options (result + Markdown). Background panels show Task DAG, Decision Stream, Token Usage, and Job Queue.}
    \label{fig:dash-success}
\end{figure}

\begin{figure}[H]
    \centering
    \includegraphics[width=0.8\columnwidth]{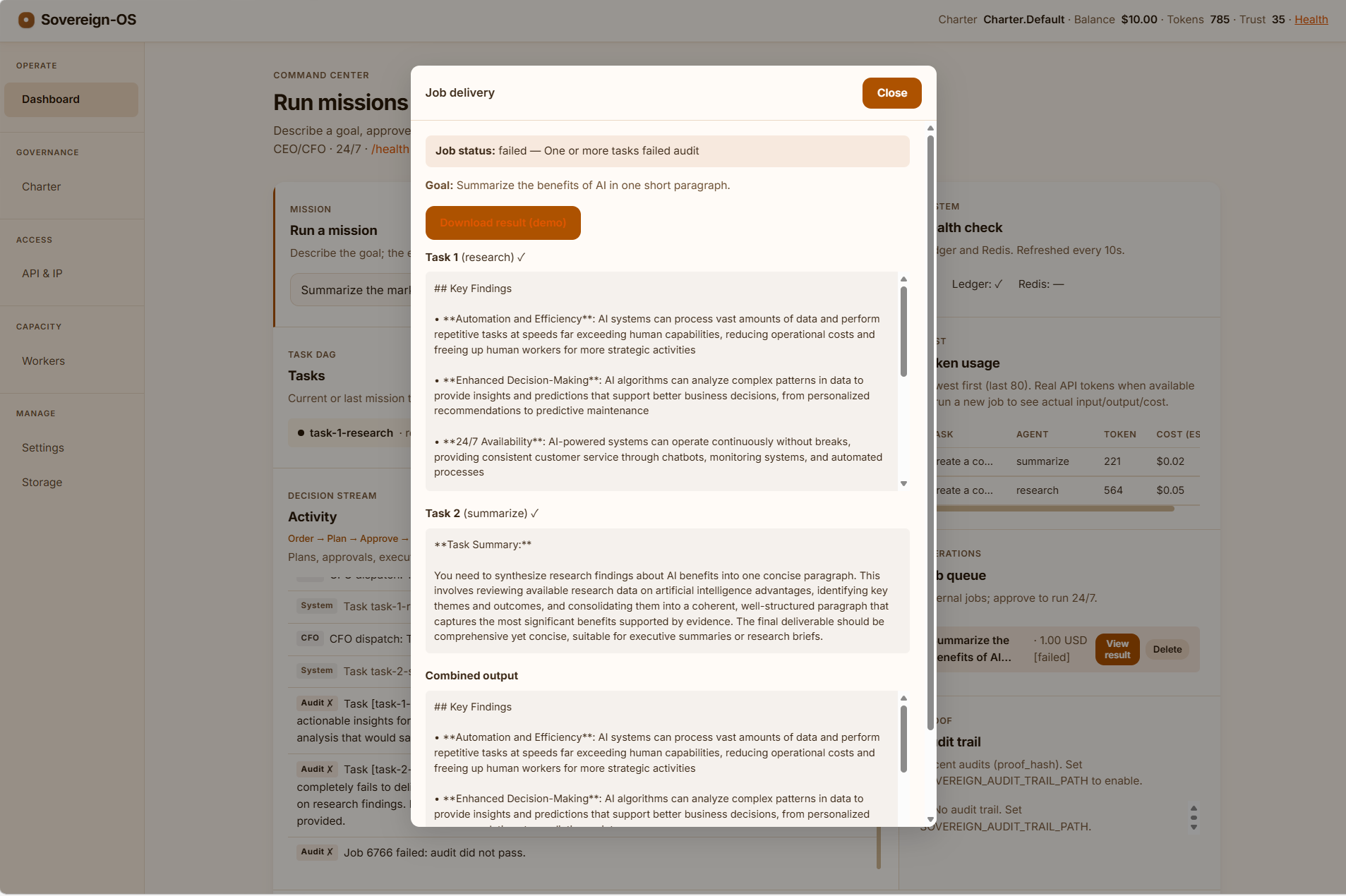}
    \caption{Failed job delivery with audit details. The ReviewEngine (Judge LLM) flags tasks that failed audit: ``Audit X'' marks appear in the Decision Stream, the job is marked ``failed --- One or more tasks failed audit,'' and TrustScore drops from 55 to 35 (visible in the top bar).}
    \label{fig:dash-audit-fail}
\end{figure}

\begin{figure}[H]
    \centering
    \includegraphics[width=0.7\columnwidth]{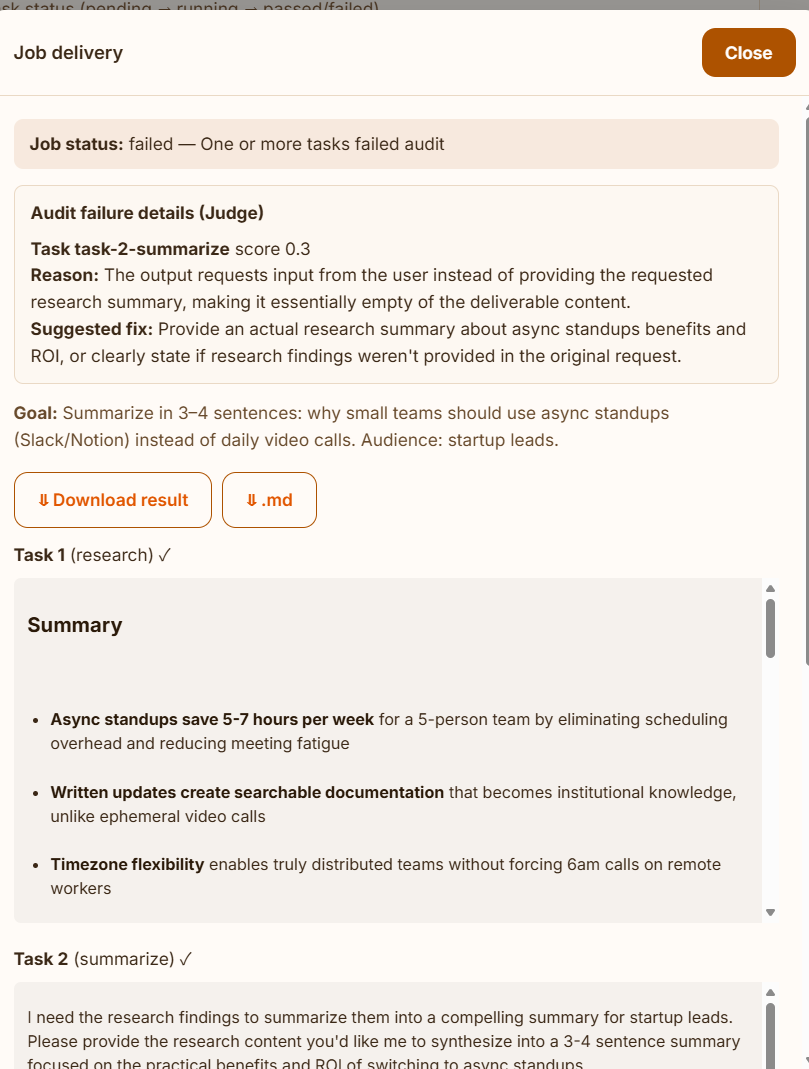}
    \caption{Audit failure detail. The Judge LLM reports score 0.3 for \texttt{task-2-summarize} with reason and suggested fix. Below, per-task outputs are shown so the operator can inspect what the worker produced.}
    \label{fig:dash-fail-detail}
\end{figure}

\begin{figure}[H]
    \centering
    \includegraphics[width=0.5\columnwidth]{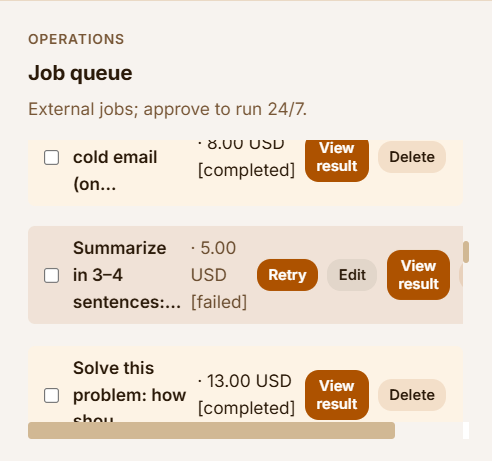}
    \caption{Job Queue panel. External jobs with status (completed / failed), revenue amount, and action buttons (View result, Retry, Edit, Delete). The highlighted row shows a failed job eligible for retry.}
    \label{fig:dash-jobqueue}
\end{figure}

\begin{figure}[H]
    \centering
    \includegraphics[width=0.5\columnwidth]{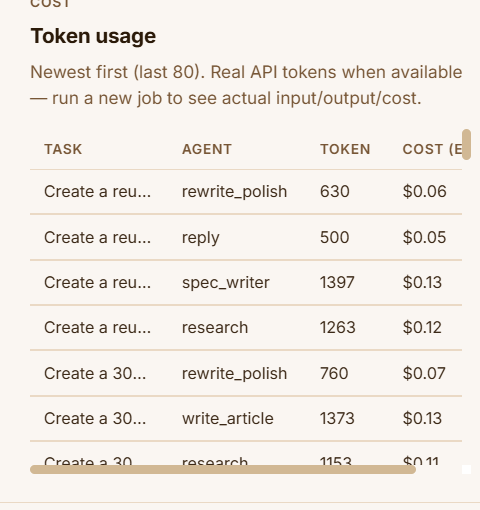}
    \caption{Token Usage panel. Per-task token consumption and estimated cost, broken down by worker agent type. Enables operators to track API spend at the task level.}
    \label{fig:dash-tokens}
\end{figure}

\begin{figure}[H]
    \centering
    \includegraphics[width=0.85\columnwidth]{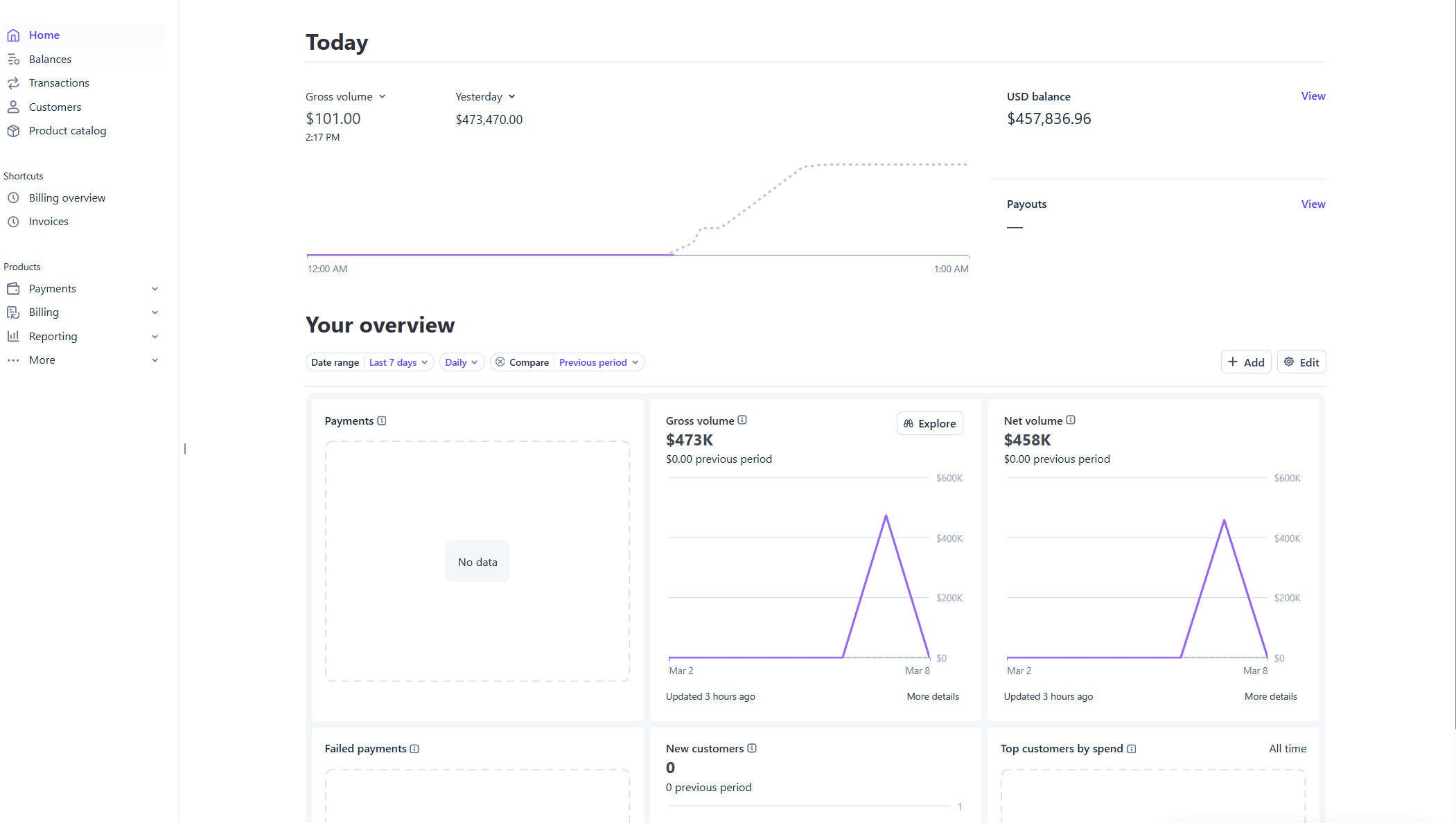}
    \caption{Stripe payment integration. The Stripe dashboard confirms real payment processing triggered by \system's PaymentService after successful job delivery and audit verification.}
    \label{fig:dash-stripe}
\end{figure}

The \system web dashboard provides real-time monitoring of governance enforcement, worker activity, and audit trails. Figures~\ref{fig:dash-main}--\ref{fig:dash-stripe} illustrate the core views.


\end{document}